\documentclass[twocolumn,tighten]{aastex62}

 \received{May 15, 2019}
\accepted{May 30, 2019}
\submitjournal{ApJL}
\shorttitle{Li-rich Kepler giants}
\shortauthors{Singh et al. 2019}
\begin{document}

\title{Survey of Li-rich giants among Kepler and LAMOST fields: Determination of Li-rich giants Evolutionary Phase}
\correspondingauthor{Raghubar Singh}
 \email{raghubar2015@gmail.com, raghubar.singh@iiap.res.in}

\author{Raghubar Singh}
 \affil{Indian Institute of Astrophysics, II Block, Koramangala, Bengaluru-560034, India }
 \affil{Pondicherry University R. V. Nagara, Kala Pet, 605014, Puducherry, India}

 \author{Bacham E. Reddy}
 \affil{Indian Institute of Astrophysics, II Block, Koramangala, Bengaluru-560034, India }

 \author{Yerra Bharat Kumar}
 \altaffiliation{LAMOST Fellow}
 \affil{Key Laboratory for Optical Astronomy, National Astronomical Observatories, Chinese Academy of Sciences, Beijing, 100101, China}

 \author{ H.M.  Antia}
 \affil{Tata Institute of Fundamental Research, Colaba- Mumbai, India}

\begin{abstract}

 In this letter, we report the discovery of 24 new super Li-rich (A(Li) $\ge$ 3.2) giants of He-core burning phase at red clump region. Results are based on systematic search of a large sample of about 12,500 giants common to the LAMOST spectroscopic and Kepler time resolved photometric surveys. The two key parameters derived from Kepler data; average period spacing ($\Delta p$) between $l=1$ mixed gravity dominated g-modes and average large frequency separation ($\Delta \nu$) $l=0$ acoustic p-modes, suggest all the Li-rich giants are in He-core burning phase. This is the first unbiased survey subjected to a robust technique of asteroseismic analysis to unambiguously determine evolutionary phase of Li-rich giants. The results provide a strong evidence that Li enhancement phenomenon is associated with giants of He-core burning phase, post He-flash, rather than any other phase on RGB with inert He-core surrounded by H-burning shell.  

\end{abstract}


\section{Introduction \label{sec:intro }}

 It is firmly established that Li-rich giants do exist, though not very common, about 1\%,  among red giants. Thanks to recent large surveys such as LAMOST \citep{Cui2012} and GALAH \citep{galah2015} there are now a few hundred Li-rich giants which have Li abundance of $A(\mathrm{Li}) \geq 1.5$~dex\footnote{$A(\mathrm{Li})= \log10(N(\mathrm{Li})/N(\mathrm{H})) + 12$}, a commonly adopted upper limit for normal giants of red giant branch \citep[RGB;][]{Iben1967a}. For example, recent large study by \cite{deepak2019} discovered more than 300 Li-rich giants from GALAH spectroscopic survey doubling the number of Li-rich giants known till then since  their first discovery  by \cite{wallerstein1982}. However, there is no consensus on the origin of Li excess in red giants which has been elusive for decades. This is because there is no clarity on their evolutionary phase, a key parameter for identifying the source of Li enhancement. 

 Presently, evolutionary phase of most of the Li-rich giants is based on their location in the $T_{\rm eff}$ -- $L$ plane of Hertzsprung-Russell (HR) diagram. Such determination is fraught with ambiguity as uncertainties in derived stellar parameters arising from different methodologies by different studies are often larger than the differences in stellar parameters of $T_{\rm eff}$ and $L$ between different locations on HR diagram. As a  result different studies suggested different phases for Li-rich giants: below the luminosity bump \citep[e.g.;][]{Casey2016}, at the bump \citep[e.g.;][]{Charbonnel2000}, red clump \citep[e.g.;][]{Kumar2011a, SilvaAguirre2014, Monaco2014, singh2019} and any where along the RGB \citep[e.g.;][]{Lebzelter2012, Martell2013}. These results led to suggestions of different scenarios for origin of Li excess in Red giants. For example, diffusion of Li upwards (in case of sub-giants), some kind of extra-mixing associated with luminosity bump (e.g.,\citealt{Palacios2001}), nucleosynthesis and dredge-up during He-flash in case of red clump \citep{Kumar2011a}, and external scenario such as mergers of planet or sub-stellar objects for occurrence of Li-rich giants any where along the RGB \citep{Lebzelter2012}. It is important to address the question whether Li-rich phenomenon is confined to a single evolutionary phase or to multiple phases on RGB. For this to be answered one would require independent method.    
 
 Asteroseismic analysis is one of the robust methods to separate giants ascending RGB with He-inert core from those with core He-burning \citep{Bedding2011} red clump giants, post He-flash. One could make use of Kepler \citep{Borucki2010} time resolved photometric data for this purpose. Unfortunately, none of the known Li-rich giants are in the Kepler fields barring a few recently reported ones. To date, there are only six Li-rich giants that were analyzed using Kepler and CoRoT asteroseismic data.  With the exemption of one \citep{Jofre2015}, all the five giants have been found to be He-core burning giants of red clump \citep{SilvaAguirre2014, Carlberg2015, kumar2018, Smiljanic2018}. It is necessary to conduct a large unbiased systematic survey of Li-rich giants which have asteroseismic data. In this letter, we show results from the survey based on  large red giant sample stars of about 12,500 that are common among LAMOST spectroscopic and Kepler  photometric surveys.   
   
\begin{figure}
\plotone{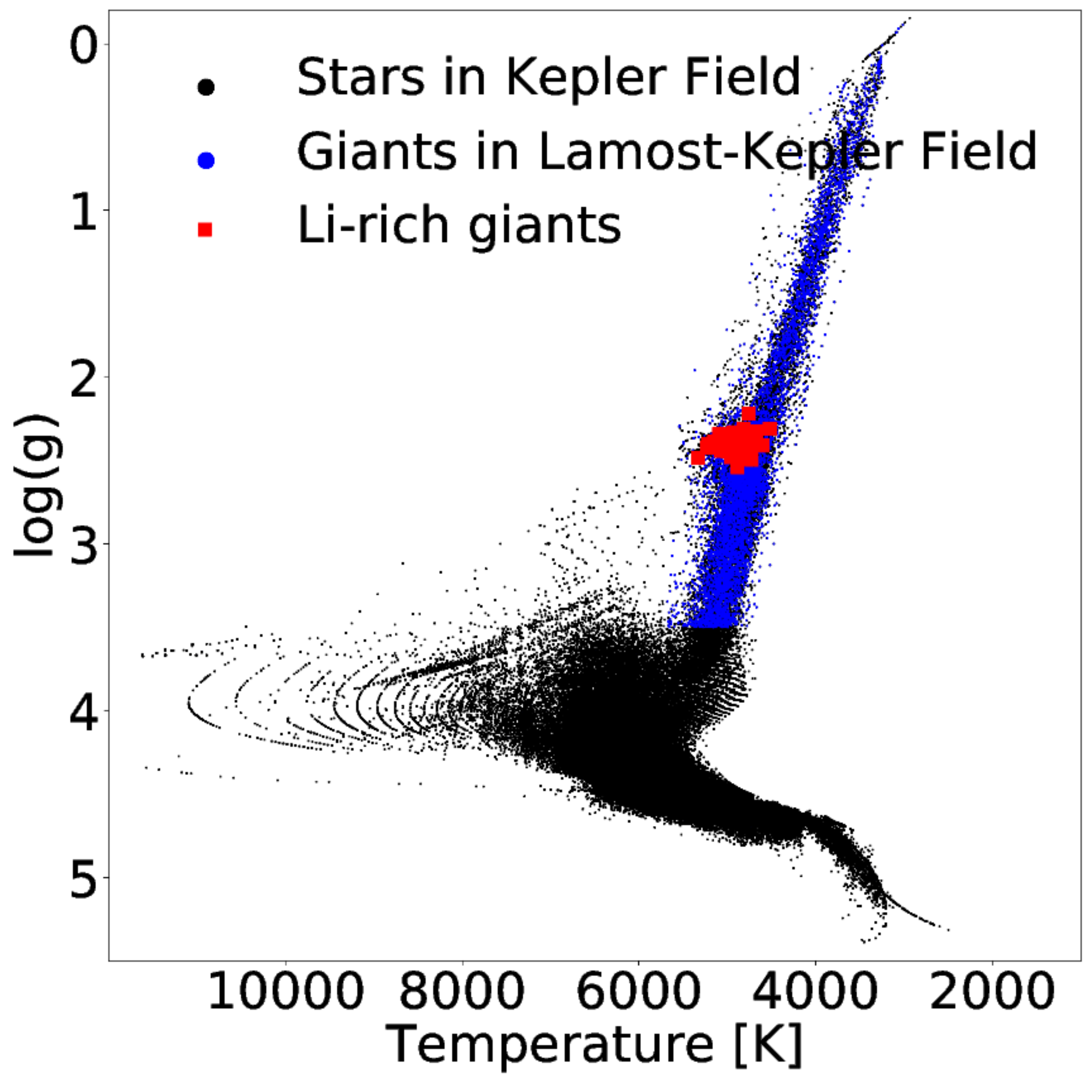}
\caption{Survey sample of 12,500 giants (Blue symbols) along with the entire sample from the Kepler catalogue as background (black symbols). Red symbols represent giants with strong Li line at 6707\AA.}
\label{fig:sample}
\end{figure}

\section{ Sample selection \label{sec:sample }}

 The primary purpose of this study is to accurately and unambiguously determine  evolutionary phase of Li-rich giants. For this we adopted a sample giants that are common among Kepler  \citep[KIC;][]{Mathur2017} and LAMOST spectroscopic catalogues. By applying criterion of $\log g \le 3.5$ and $T_\mathrm{eff} \le 5500$ K for RGB giants, we found a sample of 23,000 giants in the Kepler Input Catalogue. Of which about 12,500 giants are found to be common in LAMOST catalogue of data release 4 (DR4) \footnote{\url{http://dr4.lamost.org/}}. LAMOST is a low resolution ($R=1800$) spectroscopic survey of stars covering wavelength range of 3700 -- 9000~\AA. Continuum fitted spectra have been inspected for the presence of Li resonance line at 6707~\AA\, and found 78 spectra with strong Li line. Common sample of giants among Kepler and LAMOST (blue dots) along with the entire sample from Kepler catalogue \citep{Mathur2017} as background (black dots) are shown in Fig~\ref{fig:sample}. Stars that show strong Li line at 6707\AA\ are shown as red squares. Note that all of them are concentrated in a particular range of $\log g$ which coincides with positions of both red clump and luminosity bump in HR-diagram. 

\begin{figure*}
\plotone{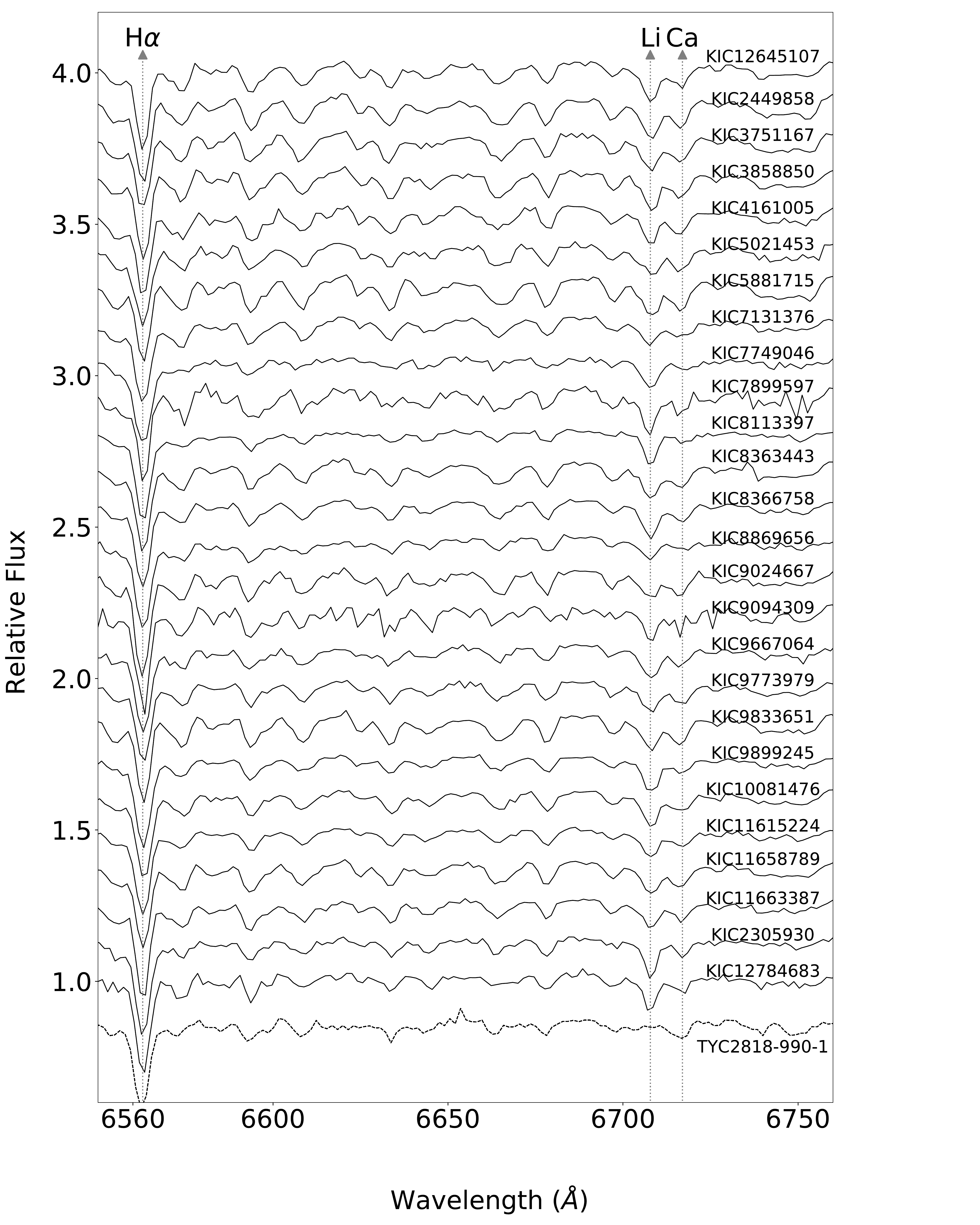}
\caption{ Spectra of 26 Li-rich giants showing exceptionally strong Li resonance line at 6707\AA. Also, shown are the two reference spectra of known super Li-rich giant (KIC~12645107, KIC~2305930) of A(Li)=3.3, 4.1 dex and a normal Li giant of A(Li) = 0.5 dex (bottom, TYC~2818-990-1).}
\label{fig:lamostspectra}
\end{figure*}

\section{ Lithium abundance \label{sec: li abundance}}

Spectra of Li-rich giants with strong Li resonance line at 6707\AA\ are shown in Fig~\ref{fig:lamostspectra} along with the known Li-rich giant KIC~12645107 on the top and a normal Li giant at the bottom. For estimating Li abundance from low resolution spectra we used a method that was successfully demonstrated previously by \cite{Kumar2011a} and \cite{bharat2018l}. This method involves measuring of Li line strength at 6707 \AA\ relative to an adjacent Ca~I line at 6717 \AA, both are zero low excitation potential lines and show similar sensitivity to $T_{\rm eff}$. The derived ratios of Li 6707 \AA\ core strength to Ca 6717 \AA\ are plugged into correlations between Li abundance and line strength ratios derived by \cite{bharat2018l}. Since uncertainties are relatively higher, 0.3 -- 0.4 dex, sample has been restricted to only giants with very strong Li line or the estimated Li abundance $A(\mathrm{Li}) \geq 3.0$~dex. This is to avoid mistaking of normal giants with $A(\mathrm{Li})$ $\leq 1.8$~dex \citep{Iben1967a}. We found 26 giants with $A(\mathrm{Li}) \geq 3.0$~dex and half a dozen have $A (\mathrm{Li}) \geq 4.0$~dex, about an order of magnitude more than the current ISM value (3.3 dex), and about a factor of 100 more than the the maximum predicted abundance of $A(\mathrm{Li}) = 1.8$~dex \citep{Iben1967a}. Estimated Li abundances along with [Fe/H], $T_{\rm eff}$ and log $g$ (stellar parameters from LAMOST DR4 catalogues) are given in Table~\ref{tab:table1}. Two of these have been recently reported as Li-rich giants \citep{kumar2018} based on high resolution spectra. Mean difference between the estimated $A(\mathrm{Li})$ in this study and the literature values based on high resolution spectra is 0.2 dex, which agrees well within uncertainties. 
 
\begin{figure}[ht!]
\plotone{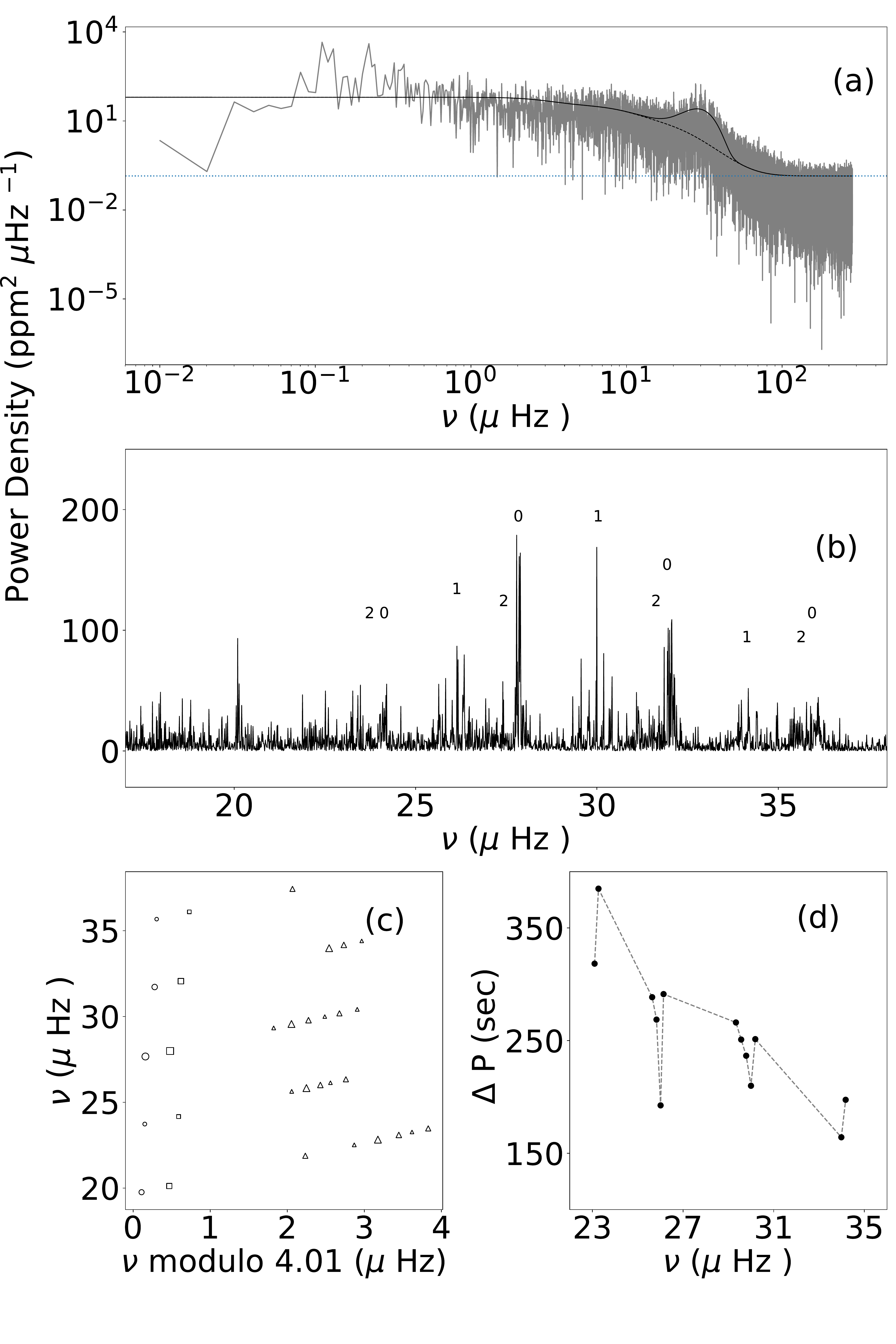}
\caption{Top panel: gray region in background is PDS of KIC~11615224 and solid black line is global background fit to the PDS. Middle panel: $l=0, 1, 2$ modes in the PDS. Bottom panel: Measurement of large frequency separation and gravity mode period spacing of star. In the bottom left panel circle are modes corresponding to $l=0$, square corresponding to $l=2$ and triangle corresponding to $l=1$ modes.  \label{fig:mod_fit}}
\end{figure}

\section{ Analysis of asteroseismic data \label{sec:astro analysis}}

 All the 26 Li-rich giants that are given in Table~\ref{tab:table1} have long cadence (29.4 min) of 10 to 17 quarters Kepler photometric data. It is known that red giants show oscillations of mixed modes of gravity (g-mode) arising from the central core and acoustics modes (p-modes) arising in the convective envelope \citep{DeRidder2009, Beck2011}. In this work we have used lightkurve package (\url{https://github.com/KeplerGO/lightkurve}) for merging individual quarters into a combined light curve, and converting the combined light curve into power density spectrum (PDS) using Lomb-Scargle Fourier transform. In Fig~\ref{fig:mod_fit}a, PDS for one of the sample stars of KIC~11615224 from Table~\ref{tab:table1} is given in which solid line is the fitting for background. Background subtracted  and smoothed PDS (Fig~\ref{fig:mod_fit}b) is used to identify modes and measuring their frequencies.  
 
 There are two key parameters: frequency separation between two consecutive radial ($l=0$) modes ($\Delta\nu$) and period separation between two consecutive dipole ($l=1$) modes ($\Delta p$) which are used to separate red giants of He-burning core of red clump and He-inert core of RGB \citep{Bedding2011}. From the smoothed PDS radial modes $l = 0 $, dipole modes ($l=1$) and quadruple modes ($l=2$) have been identified and the corresponding frequencies for 4--5 modes ($l=0$) in each star's PDS  are measured. Values of $\Delta\nu$ are those for which modulo or remainder of $\nu/\Delta\nu$ is same for  the frequencies of respective measured mode. In Fig~\ref{fig:mod_fit}c, this has been illustrated for a typical giant, KIC~1165224. For its modes of $l=0$, $l=2$ and $l=1$, we found a value of $\Delta\nu =$ 4.01 $\mu$Hz for which modulo is same for all the frequencies. The values of derived large frequency separation, $\Delta\nu$ are given in Table~\ref{tab:table1}. After measuring frequency corresponding to detected dipole modes ($l=1$, g-modes), we calculate the period of dipole modes and period spacing between the consecutive dipole modes which are given in Fig~\ref{fig:mod_fit}d. Median value of derived period spacing is considered as gravity mode period spacing of a star \citep{Stello2013}. Derived values of $\Delta\nu$ and $\Delta p$, given in Table~\ref{tab:table1}, suggest all the Li-rich giants are He-core burning red clump stars. Our analysis agrees well with the recent study done by \cite{Yu2018}. Average difference between ours and \cite{Yu2018} for $\Delta \nu$ and $\nu_{max}$  is 0.8 and 0.1 $\mu$Hz respectively. 

  Stellar parameters such as radius and mass have been derived using the seismic parameters and the calibrations are given by \citep{Bedding1995}. All the giants are low mass with M $\leq$ 2.0 M$_{\odot}$ (Table~\ref{tab:table1}). Further, light curves show no indication of a major flaring activity which is in agreement with lack of visible asymmetry or emission profiles of $H_\alpha$ in the spectra (Fig~\ref{fig:lamostspectra}) indicating no significant stellar activity in the photospheres of the stars. 

\begin{deluxetable*}{ccccccccccccc}
\tablecaption{List of Li-rich giants \label{tab:table1}}
\tablehead{
\colhead{KIC} & \colhead{Vmag} & \colhead{$A(\mathrm{Li})$\tablenotemark{a}} & \colhead{$\Delta P$ } & \colhead{A(Li)} & \colhead{$\nu_{max} $}& \colhead{$\Delta \nu$} & \colhead{$T_{\rm eff}$} & \colhead{mass} & \colhead{radius}& \colhead{$\log g$} & \colhead{[Fe/H]} & \colhead{M}\tablenotemark{m}\\
}
\colnumbers
\decimals
\startdata
2305930  & 11.02 & 4.20 & 226 & 4.1 & 27.2  & 4.0  & $ 4861 \pm 40 $  & $ 0.71  \pm 0.1$  & $ 9.40 \pm 0.7 $ & $ 2.40 \pm 0.1 $  & $-0.50  \pm 0.0$   & 2 \\
2449858  & 13.38 & ---  & 235 & 3.3 & 26.8  & 3.5  & $ 4840 \pm 30 $  & $ 1.15  \pm 0.1$  & $ 12.1 \pm 0.5 $ & $ 2.50 \pm 0.1 $  & $-0.15  \pm 0.0$   & 1 \\
3751167  & 13.83 & ---  & 419 & 4.0 & 26.1  & 3.6  & $ 4777 \pm 239 $ & $ 0.91  \pm 0.2$  & $ 10.9 \pm 1.0 $ & $ 2.25 \pm 0.4 $  & $-1.06  \pm 0.2$   & 2 \\
3858850  & 12.44 & ---  & 192 & 3.3 & 25.9  & 3.5  & $ 4434 \pm 50 $  & $ 0.90  \pm 0.1$  & $ 11.1 \pm 0.6 $ & $ 2.62 \pm 0.1 $  & $ 0.27  \pm 0.1$   & 1 \\
4161005  & 13.93 & ---  & 247 & 3.3 & 29.1  & 3.9  & $ 4897 \pm 40 $  & $ 0.93  \pm 0.2$  & $ 10.7 \pm 0.8 $ & $ 2.35 \pm 0.1 $  & $-0.52  \pm 0.0$   & 1 \\
5021453  & 11.25 & ---  & 314 & 4.0 & 31.8  & 4.0  & $ 4754 \pm 26 $  & $ 1.02  \pm 0.1$  & $ 10.5 \pm 0.6 $ & $ 2.55 \pm 0.1 $  & $-0.08  \pm 0.0$   & 1 \\
5881715  & 11.64 & ---  & 191 & 3.8 & 30.9  & 3.4  & $ 4786 \pm 35 $  & $ 1.85  \pm 0.2$  & $ 14.2 \pm 0.7 $ & $ 2.35 \pm 0.1 $  & $-0.15  \pm 0.0$   & 1 \\
7131376  & 13.99 & ---  & 190 & 3.8 & 34.8  & 4.1  & $ 4696 \pm 80 $  & $ 1.18  \pm 0.1$  & $ 10.8 \pm 0.4 $ & $ 2.68 \pm 0.1 $  & $ 0.08  \pm 0.1$   & 1 \\
7749046  & 13.47 & ---  & 235 & 4.2 & 29.9  & 3.8  & $ 4891 \pm 26 $  & $ 1.13  \pm 0.1$  & $ 11.2 \pm 0.4 $ & $ 2.36 \pm 0.1 $  & $-0.71  \pm 0.0$   & 3 \\
7899597  & 13.61 & ---  & 224 & 3.9 & 31.6  & 3.8  & $ 4710 \pm 50 $  & $ 1.26  \pm 0.2$  & $ 11.7 \pm 0.9 $ & $ 2.49 \pm 0.1 $  & $-0.10  \pm 0.1$   & 1 \\
8113379  & 13.13 & ---  & 273 & 3.2 & 31.2  & 3.9  & $ 4757 \pm 40 $  & $ 1.06  \pm 0.1$  & $ 10.7 \pm 0.4 $ & $ 2.53 \pm 0.1 $  & $-0.06  \pm 0.0$   & 1 \\
8363443  & 10.95 & ---  & 217 & 3.5 & 32.4  & 3.8  & $ 4490 \pm 40 $  & $ 1.28  \pm 0.1$  & $ 11.8 \pm 0.3 $ & $ 2.58 \pm 0.1 $  & $ 0.23  \pm 0.0$   & 1 \\
8366758  & 12.50 & ---  & 218 & 3.8 & 26.4  & 3.9  & $ 4664 \pm 50 $  & $ 0.67  \pm 0.1$  & $ 9.3  \pm 0.3 $ & $ 2.57 \pm 0.1 $  & $ 0.18  \pm 0.0$   & 1 \\
8869656  & 9.34  & ---  & 240 & 4.1 & 30.7  & 3.8  & $ 4764 \pm 30 $  & $ 1.15  \pm 0.1$  & $ 11.3 \pm 0.5 $ & $ 2.44 \pm 0.1 $  & $-0.30  \pm 0.0$   & 1 \\
9024667  & 12.28 & ---  & 449 & 3.4 & 25.2  & 3.5  & $ 4555 \pm 35 $  & $ 0.83  \pm 0.1$  & $ 10.7 \pm 0.7 $ & $ 2.59 \pm 0.1 $  & $ 0.16  \pm 0.0$   & 1 \\
9094309  & 14.31 & ---  & 253 & 4.0 & 33.2  & 4.1  & $ 4919 \pm 129 $ & $ 1.16  \pm 0.1$  & $ 10.8 \pm 0.5 $ & $ 2.55 \pm 0.2 $  & $-0.32  \pm 0.1$   & 2 \\
9667064  & 13.35 & ---  & 176 & 4.4 & 30.2  & 3.6  & $ 4678 \pm 211 $ & $ 1.34  \pm 0.2$  & $ 12.4 \pm 0.7 $ & $ 2.28 \pm 0.3 $  & $-0.10  \pm 0.2$   & 1 \\
9773979  & 14.30 & ---  & 487 & 3.2 & 32.6  & 4.0  & $ 4622 \pm 86 $  & $ 1.15  \pm 0.2$  & $ 11.0 \pm 0.6 $ & $ 2.48 \pm 0.1 $  & $-0.05  \pm 0.1$   & 2 \\
9833651  & 12.52 & ---  & 174 & 3.6 & 38.8  & 4.2  & $ 4683 \pm 44 $  & $ 1.47  \pm 0.1$  & $ 11.4 \pm 0.4 $ & $ 2.67 \pm 0.1 $  & $ 0.09  \pm 0.0$   & 1 \\
9899245  & 13.04 & ---  & 150 & 3.4 & 33.2  & 3.9  & $ 4700 \pm 30 $  & $ 1.30  \pm 0.3$  & $ 11.6 \pm 0.8 $ & $ 2.72 \pm 0.1 $  & $ 0.10  \pm 0.0$   & 1 \\
10081476 & 13.84 & ---  & 211 & 3.8 & 26.6  & 3.4  & $ 4453 \pm 50 $  & $ 1.07  \pm 0.2$  & $ 11.9 \pm 0.8 $ & $ 2.52 \pm 0.1 $  & $ 0.24  \pm 0.0$   & 1 \\
11615224 & 11.16 & ---  & 257 & 3.3 & 30.0  & 4.0  & $ 4746 \pm 25 $  & $ 0.85  \pm 0.1$  & $ 9.80 \pm 0.4 $ & $ 2.40 \pm 0.1 $ & $-0.04  \pm 0.02$  & 2 \\
11658789 & 13.36 & ---  & 228 & 3.9 & 31.2  & 4.3  & $ 4999 \pm 75 $  & $ 0.81  \pm 0.1$  & $ 9.30 \pm 0.6 $ & $ 2.48 \pm 0.1 $  & $-0.70  \pm 0.1$   & 2 \\
11663387 & 12.59 & ---  & 217 & 4.0 & 32.7  & 4.1  & $ 4642 \pm 40 $  & $ 1.01  \pm 0.1$  & $ 10.4 \pm 0.2 $ & $ 2.49 \pm 0.1 $  & $ 0.02  \pm 0.0$   & 1 \\
12645107 & 11.40 & 3.24  & 243 & 3.5 & 30.4  & 3.8  & $ 4853 \pm 40 $  & $ 1.14  \pm 0.1$  & $ 11.2 \pm 0.4 $ & $ 2.39 \pm 0.1 $  & $-0.22  \pm 0.0$   & 1 \\
12784683 & 11.10 & ---  & 239 & 3.4 & 28.7  & 3.7  & $ 4862 \pm 25 $  & $ 1.11  \pm 0.2$  & $ 11.4 \pm 0.7 $ & $ 2.33 \pm 0.1 $  & $-0.28  \pm 0.0$   & 1 \\
\enddata
\tablenotetext{a}{Li abundance of two stars from \cite{kumar2018}}
\tablenotetext{m}{  Milky way membership; 1:Thin disk, 2:Thick disk, 3:Halo star}
\tablecomments{All are red clump star in classification of  \cite{Yu2018} also.}
\end{deluxetable*}

\section{ Discussion \label{sec:discussion}}

\begin{figure}[ht!]
\plotone{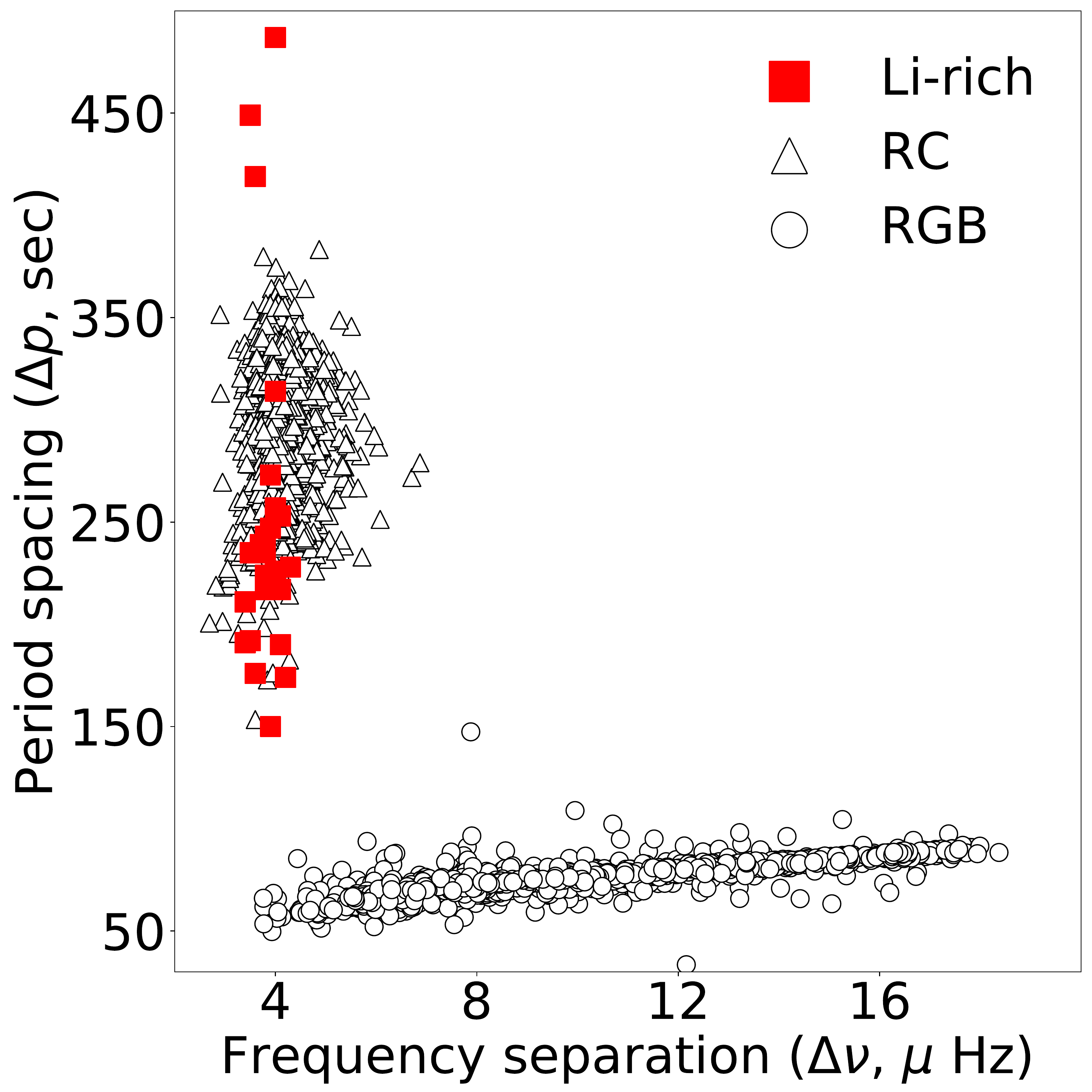}
\caption{Li-rich giants discovered in this study (red squares) are shown in $\Delta\nu - \Delta p$ asteroseismic diagram. Giants classified based on asteroseismic analysis form the background: He-core burning RC giants (open triangle) and inert He-core giants ascending RGB first time (open circle). Note, all the Li-rich giants fall in the RC region of the diagram.}
\label{fig:ast_hrd}
\end{figure}

This is the first of its kind survey based on  a large unbiased sample survey of giants that are common among LAMOST spectroscopic and Kepler photometric surveys. Analysis yielded a total of 26 Li-rich giants.  Abundance results along with the derived asteroseismic parameters; $\Delta\nu$ and $\Delta p$ are given in Table~\ref{tab:table1}, and are shown in the  plot of $\Delta p$ and $\Delta \nu$ (Fig~\ref{fig:ast_hrd}). Known RC and RGB giants based on asteroseismic analysis form background. As shown in Figure~\ref{fig:ast_hrd}, all the Li-rich giants from this study show a large values of $\Delta p\geq 150$ sec and small values of $\Delta \nu\leq$ 5 $\mu$Hz, and occupy the He-core burning phase region of $\Delta\nu - \Delta p$ diagram \citep{Bedding2011}. Interestingly, none of the Li-rich giants found in this survey are on ascending RGB. Results imply Li enhancement phenomenon is most probably associated with He-flash at the tip of RGB or post He-flash rather than on RGB. 

 As earlier stated,  there are only six Li-rich giants for which evolutionary phase is determined based on asteroseismology. All of them were discovered serendipitously. Of which five are unambiguously classified as giants of He-core burning phase in red clump region \citep{SilvaAguirre2014, Carlberg2015, kumar2018, Smiljanic2018} and one as RGB near the bump\citep{Jofre2015}. The lone exception being the KIC~9821622 \citep{Jofre2015}. Though this is a bonafide RGB star with inert He-core and H-burning shell ($\Delta\nu=$6.07 $\mu$Hz and $\Delta p= 67.6$ sec) its categorization as Li-rich giant is not beyond doubt as the Li abundance (LTE: $A(\mathrm{Li}) = 1.49$~dex and NLTE: 1.65~dex) measured from well defined and much stronger line at 6707\AA\ is at the borderline. It is important to establish whether this particular star is indeed a Li-rich giant as it has serious implications for identifying source of Li enrichment in red giants. 
 
 Of course, there are many Li-rich giants in literature which were reported as being on RGB based on their positions in the HR-diagram. This method found to be uncertain in determining exact evolutionary phase. For example, of the five asteroseismically known Li-rich RC giants, KIC~4937011 was initially reported as RGB star below the bump by \cite{Anthonytwarog2013,Carlberg2015} based on its location in HR diagram. However, its derived asteroseismic parameters $\Delta\nu=$ 4.15 $\mu$Hz and $\Delta p = 249.9$ sec \citep{Vrard2016} firmly puts the  giant in He-core burning phase. This illustrates the difficulty of determining their precise evolutionary phase. We also note another recent study by \cite{Yan2018} in which they reported TYC 429-2097-1 as the most Li-rich giant with $A(\mathrm{Li}) = 4.51$~dex and located at the bump based on their location in HR-diagram. This star is not in the Kepler field. However, its derived ratio of [C/N] = $-0.47 \pm 0.10$  is more compatible with it being in red clump \citep{Hawkins2016, singh2019}. In fact, \cite{Casey2016} made an important observation that although the stellar parameters of majority of Li-rich giants is consistent with being on RGB at or below the luminosity bump, they are each individually consistent with being RC. However, they couldn't conclude that they are indeed RC giants based on available data to them. Does this mean, a number of Li-rich giants that are reported to be on RGB based on $L$ and $T_{\rm eff}$ are  misclassified?  Results in this study seem to suggest this is a real possibility.

 Due to ambiguity in their evolutionary phase, numerous models have been constructed to explain Li excess with prevailing conditions at each of the suggested multiple phases on RGB. Broadly, theoretical models fall into two categories: external and in-situ scenarios. One of the external scenarios is merger of planets or sub-stellar objects such as brown dwarfs. This is invoked with the expectation that planet or brown dwarfs contain reservoir of primordial Li with little/no depletion, and their mergers will enhance star's photospheric Li abundance either by direct addition of Li reservoir to the photosphere or by induced mixing due to angular momentum transfer to the star or a combination of the both \citep[see e.g.;][]{Siess1999, Casey2016}. This scenario gained merit with the evidence that the large planets in close-in orbits among sub-giants are less frequent compared to their counter parts on main sequence stars \citep{Villaver2014}.
 
However, to account for levels of Li seen in many of the super Li-rich giants, one would require merger of several Jupiter size planets having undiluted Li reservoir \citep{Carlberg2012}. Merger of such large number of planets is very unlikely. Also, theoretical models put a maximum limit of Li abundance due to such mergers at $A(\mathrm{Li}) = 2.2$~dex \citep{Aguilera2016}.  Further, contrary to our results engulfment scenario suggests occurrence of Li-rich giants anywhere along the RGB and presence of infrared excess as a result of merger impact. Results support the argument by \cite{deepak2019} against external scenario based on the frequency of Li-rich giants occurring at various phases. They show that disproportionately large number of Li-rich giants belong to red clump compared to any other phase on RGB. 

 Second scenario is in-situ origin which is  Li production via Cameron \& Fowler mechanism \citep{Cameron1971}, $^{3}\mathrm{He}(\alpha,\gamma)^{7}\mathrm{Be}(e^{-}, \nu)^{7}$Li. In case of Li enhancement in the photospheres of highly evolved massiv ($\geq 3-4M_{\odot}$) asymptotic giants branch (AGB) stars \citep{Lambert1989} this mechanism is expected to operate just below the convective envelope which is hot enough to produce $^{7}$Be and close enough for $^{7}$ Be to get transported to cooler upper layers where it can form $^{7}$Li. This process is known as Hot Bottom burning (HBB). In case of low mass RGB giants convection between the H-burning shell and the outer convective layers is inhibited by the radiative zone, and standard models \citep{Iben1967a} do not predict changes in abundances post 1st dredge-up. However, observations of giants post 1st dredge-up do show severe depletion of Li and reduction in the 12C/13C ratios \citep{Gilroy1991} compared to standard models of 1st dredge-up on RGB \cite{Iben1967a}. For example, severe depletion of Li starting from the bump has been well illustrated for giants in globular cluster NGC~6397 by \cite{Lind2009b}. These anomalies were explained by extra mixing at the luminosity bump at which the barrier for the deep mixing is erased \citep[see for example][]{Eggleton2008}. Ironically, bump has also been suggested as a source of Li enhancement \citep{Palacios2001,Charbonnel2005, Denissenkov2009} as many early observations showed Li-rich giants coinciding with the bump in the HR-diagram \citep{Charbonnel2000}. It would be a challenging task for explaining Li enhancement at the same phase where severe Li depletion is known to occur. Even if we assume bump as the origin for Li excess, it is unlikely Li can survive through deep convection phase from bump to the clump, and importantly, sustaining high levels of Li abundances seen in many of the super Li-rich giants. Given the Li-rich phase of RGB is a transient phenomenon lasting for a short period of about 6M years \citep{Palacios2001} compared to, for example, 50--100M years (see also \cite{deepak2019}) long evolutionary period from the bump to the RGB tip, it is very unlikely the high Li abundances seen in these red clump stars originated at the luminosity bump.      

\section{Conclusion \label{sec:conclusion}}

 In this study we addressed one of the long standing problems of precisely determining stellar evolutionary phase of Li-rich giants. Results are based on a large unbiased sample of 12,500 low mass RGB giants common among Kepler photometric and LAMOST spectroscopic surveys. We found 24 new Li-rich giants with Li abundance of $A(\mathrm{Li}) \geq 3.0$~dex, more than an order of magnitude larger compared to the maximum predicted abundance of $A(\mathrm{Li}) = 1.80$~dex. Importantly, the derived  asteroseismic parameters; $\Delta \nu$ and $\Delta p$  show all the Li-rich giants are in He-core burning phase, and are at red clump region. This is the most unambiguous evidence so far suggesting that the Li enhancement phenomenon is probably associated only with He-core burning phase rather that on RGB with inert He-core. He-flash at the tip, an immediate preceding event to red clump, may be explored for Li excess in red clump giants.

\section*{Acknowledgement}

Y.B.K thanks the support from NSFC grant no. 11850410437. Funding for LAMOST (\url{www.lamost.org}) has been provided by the Chinese NDRC. LAMOST is operated and managed by the National Astronomical Observatories, CAS. We thank entire Kepler team without whom this work could not have been possible. We also thank anonymous referee for the constructive suggestions for content improvement. 

\bibliographystyle{aasjournal}

\end{document}